# Concentration-adjustable micromixer using droplet injection into a microchannel

Ryosuke Sakurai[1,+], Ken Yamamoto[1,2,*,+], Masahiro Motosuke[1,2]

[1]Department of Mechanical Engineering, Tokyo University of Science, 6-3-1 Niijuku, Katsushika-ku, Tokyo 125-8585, Japan

[2]Research Institute for Science and Technology, Tokyo University of Science, 6-3-1 Niijuku, Katsushika-ku, Tokyo 125-8585, Japan

[*]Corresponding author, yam@rs.tus.ac.jp

[+]these authors contributed equally to this work

**Abstract**

A novel micromixing technique that exploit a thrust of droplets into the mixing interface is developed. The technique enhances the mixing by injecting immiscible droplets in a mixing channel and the methodology enables a control of the mixing level simply by changing the droplet injection frequency. We experimentally characterize the mixing performance with various droplet injection frequencies, channel geometries, and diffusion coefficients. Consequently, it is revealed that the mixing level increases with the injection frequency, the droplet-diameter-to-channel-width ratio, and the diffusion coefficient. Moreover, the mixing level is found to be a linear function of the droplet volume fraction in the mixing section. The results suggest that the developed technique can produce a large amount of sample solution whose concentration is arbitrary and precisely controllable with a simple and stable operation.

**Introduction**

Mixing of biological or chemical species is one of the essential processes in lab-on-a-chip and µTAS devices. Above all, the performance of the devices for such as the digital polymerase chain reaction (dPCR)[1,2], single-molecule analysis[3], chemical synthesis[4], or crystallization[5] is markedly dependent on the concentration control accuracy and the solution homogeneity. To achieve an effective, rapid, and complete mixing, various micromixers were developed to enhance the mixing in microscale that is typically dominated by diffusion.

The enhancement of the mixing is accomplished by increasing the area of mixing front. Micromixer is categorized into the active and the passive types by their mixing enhancement methods[6], and the active ones increase the mixing interface by generating disorder with an aid of the applied external energy. Electrokinetic flows such as the alternating-current electroosmosis (ACEO)[7] and electrothermal (ET) flows[8], temperature rise to raise the diffusion coefficient[9,10], or surface acoustic waves (SAWs)[11–13] are typical examples of the active mixer. The active mixing has an advantage in the on-demand controllability of the mixing characteristics by adjusting the applied external energy. On the other hand, the devices tend to be complicated because of extra equipments such as external power supply. Passive micromixers create new mixing interface[14] by designing mixing channels to have complex and three-dimensional microstructures[15–19], or to generate chaotic flows[20,21]. Apart from the "classical" passive mixers, some devices introduce droplets or bubbles to generate the disorder of the mixing interface in more dynamic manner[22–29]. Although most of the passive mixers cannot modify the mixing characteristics, their packages can be relatively compact and simple as they do not require the additional external sources.

Although the passive mixers generally have difficulty in the on-demand concentration control as discussed above, they are considered to be prospects because of their cost-effective and simple character. To obtain a wide range of concentration in a passive system, gradient generators[30–33], which generate serial chemical gradients by tree-like flow paths, and dilutors[34–36], which generate a concentration gradient in specific areas and extract the targeted solution, are often employed. However, they are relatively time-consuming due to the slow diffusion in microscale and require a precise chip fabrication and handling. Because these shortcomings come from the diffusion-dominant and pressure-sensitive nature of the devices, we can expect a high-performance device by installing a diffusion-free mixing technique with less branches while keeping the whole system simple. Among various passive systems, multiphase-flow-type mixers can generate large disorder by introducing immiscible droplets or bubbles into the mixing channel[14, 37]. Although they require the addition of the immiscible fluid, they have large advantages in keeping the device design simple and thus in being integrated into composite analysis system with simple fabrication technique. Moreover, the technique has less influence on biological samples in comparison with the techniques using electrical or thermal effects[14]. Garstecki *et al.*[23] introduced bubbles into a series of branching (repeatedly split and recombined) microchannels and achieved an effective mixing by fluctuating pressure between the branches. Günther *et al.*[24] utilized bubbles to form a slug (Taylor) flow in a curved microchannel to divide the sample flow into small segment in which the diffusion length can be shorten. Mao *et al.*[25] implemented chambers in a microchannel downstream of a flow-focusing device where bubbles are generated and the mixing was enhanced by a chaotic flow induced by random movement of the bubbles in the chambers.

While aforementioned mixers aim to obtain the complete mixing (*i.e.*, homogeneity of the solution), mixing *inside* the droplets are often exploited to control the concentration of the solution[34]. However, they typically require sequential and repetitive procedures to obtain the demanded concentration and therefore the operation tends to be complicated. To achieve the concentration controllability and the mixing enhancement at the same time, a multiphase-flow type mixer, which control the concentration *outside* the droplets to provide high throughput and unlimited production volume is developed. The device enhances the mixing by generating a disorder on the mixing interface and the concentration adjustment are achieved by controlling the level of the generated disorder. The disorder is generated by striking the solution–buffer interface with immiscible droplets, and the disorder level is adjusted by varying the droplet frequency (droplets per second). Furthermore, effects of the channel geometry and the diffusion coefficient of the sample are also investigated. The developed device successfully shows fine concentration controllability and high responsiveness with simple and stable operation.

## Results

**Mixing enhancement by the disorder**

A schematic of the micromixing device, which effectively mix solutions by means of the droplet injection, is shown in Fig. 1(a). All the channel height is set to 50 μm. The droplets are generated at a T-junction where a continuous-phase channel (100 μm wide) and a dispersed-phase channel (50 μm wide) confluent [Figs. 1(b) and 1(c)]. In the downstream region, the continuous phase works as a buffer and is mixed with sample solution in a mixing channel. The continuous-phase channel and the sample-solution channel (100 μm wide) confluent with an angle $\theta_m$ [Fig. 1(d)] and the buffer (including droplets) and the sample flow in the mixing channel whose width and length are $W_m$ and 10 mm, respectively. At the confluence point, the sample–buffer interface is perturbed, which results in enhancing the mixing efficiency, as the droplets carried by the continuous phase strike the interface. At the end of the

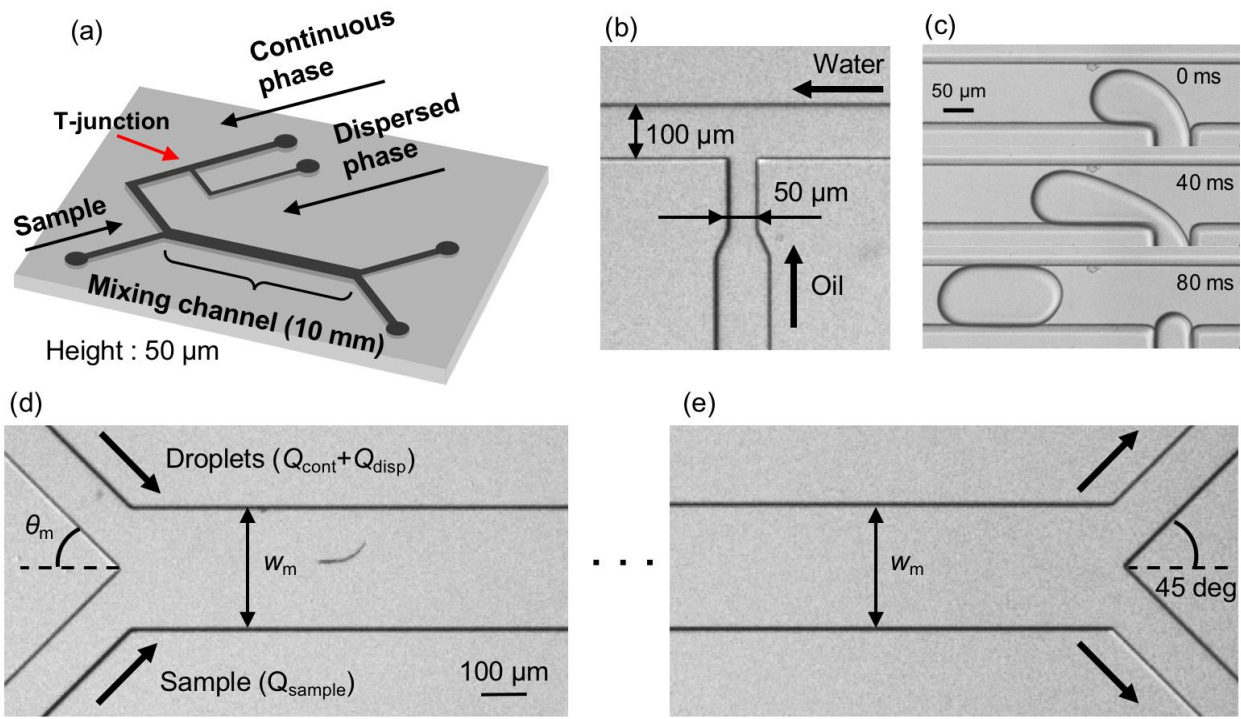

Fig. 1 (a) Overview of the micromixer exploiting a disorder induced by the droplet injection. Mixer consists of three inlets, which are used for inflow of the continuous phase (water), the dispersed phase (oil), and the sample, a T-junction for a droplet generation, a mixing channel (length of 10 mm), and two outlets. (b) and (c) Droplets are generated at the T-junction and (d) are injected to the confluence point. The mixing is enhanced by the injected droplets that induce the disorder on the interface of the continuous phase and the sample. (e) The concentration-adjusted sample is extracted from the upper-side outlet.

mixing channel, the mixed sample is extracted from an upper-side bifurcation channel [100 μm wide, Fig. 1(e)]. DI water and oleic acid are employed as the continuous phase and the dispersed phase, respectively, and either dye solution or particle solution is employed as the sample phase.

The mixing performance is evaluated with a constant flow rate of 420 μL h$^{-1}$ for the sample flow ($Q_{sample}$) and the continuous phase (including the droplets) flow ($Q_{cont} + Q_{disp}$). The frequency of the droplet generation $f$ is controlled by changing a flow-rate ratio of the dispersed phase to the continuous phase ($\varphi = Q_{disp} / Q_{cont}$), while keeping the sum of the flow rate of 420 μL h$^{-1}$. Figure 2 shows the mixing of the dye solution and the buffer. In a case of the flow without the droplets ($Q_{cont} = 420$ μL h$^{-1}$, $Q_{disp} = 0$ μL h$^{-1}$) diffusion is dominant in the mixing as shown in Fig. 2(a). On the other hand, the dominant mechanism changes by introducing droplets [Fig. 2(b)]. In this case, the immiscible droplets perturb the mixing front and the mixing enhancement is clearly visible at the end of the mixing channel. Successive images of Fig. 2(c) indicate that the disorder of the interface is induced when the droplets are injected into the mixing channel. The striking motion of the droplets displaces a certain amount of the sample towards the buffer solution region. It implies that the concentration of the mixed solution is adjustable by controlling the droplet injection. Moreover, because the droplets confine the outlet channel, they also work as a homogenizer of the mixed (but not spatially homogenized) solution by forming a Taylor flow in the outlet channel in which solutions are rapidly homogenized due to the circulated flow inside the slugs[24]. In the following sections, we investigate effects of the droplet injection on the mixing in detail by measuring the local mixing level of five cross sections ($x_m$ = 0, 2.5, 5.0, 7.5, and 10 mm downstream of the confluence point.) The mixing level is quantitatively evaluated by the relative mixing index (RMI)[38], which is the standard deviation of the light intensity distribution, in a region indicated with a red-dashed rectangle in Fig. 2(a). The RMI indicates the mixing level with a value between 1 (two phases are completely separated) and 0 (two phases are completely mixed and homogenized).

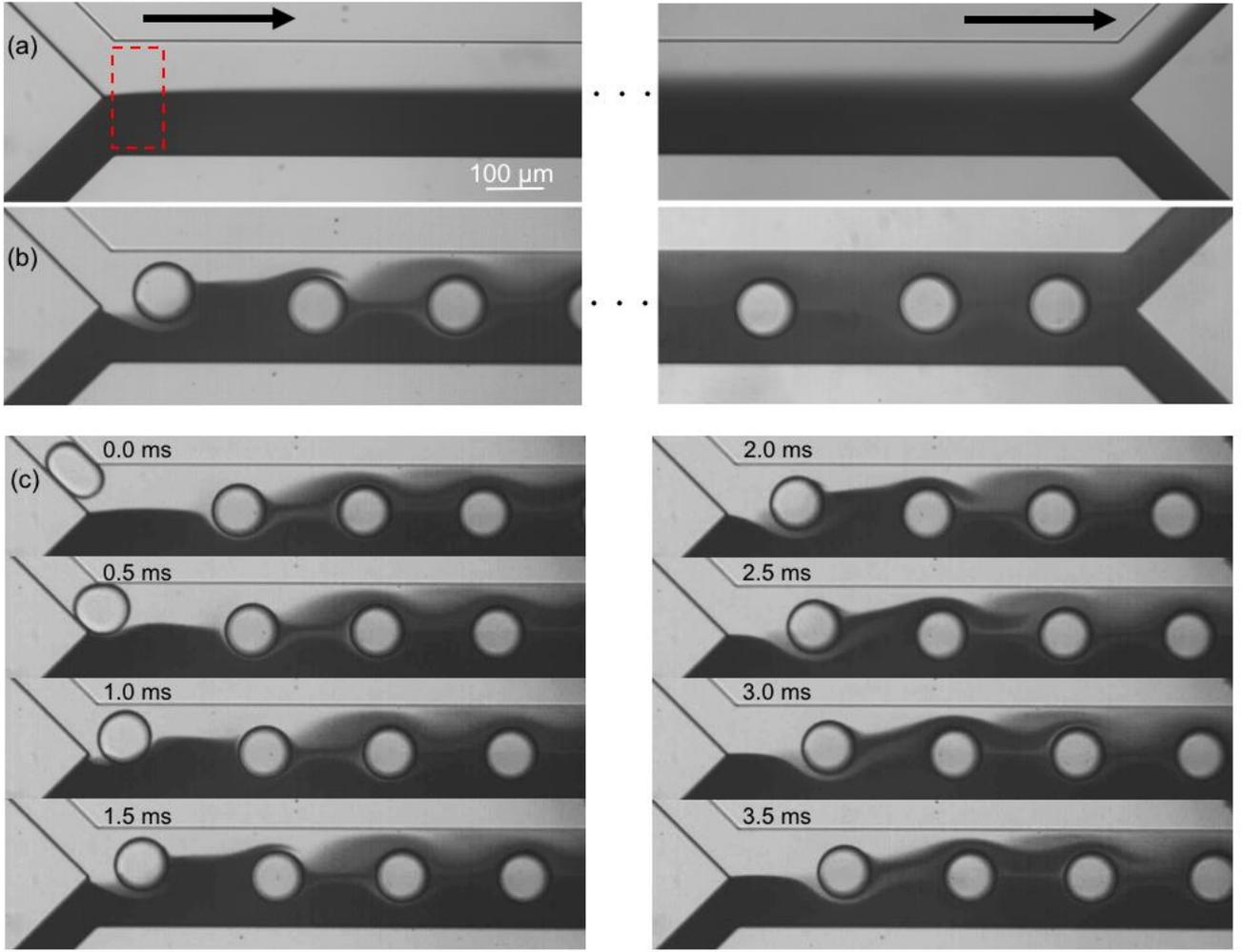

Fig. 2 (a) and (b) Mixing of two fluids at the inlet (left panels) and the outlet (right panels) of the mixing channel. (a) Diffusive mixing is dominant on the interface of the continuous phase and the sample in the case of no droplet injection, whereas (b) mixing is enhanced at the bifurcation point because of a disorder generated by the droplet injection to the interface. Red box in (a) is one of a regions for the RMI analysis. (c) Successive images of the disorder generation. The droplet shape changes from the slag to the spherical as the droplets are injected into the confluence point where the channel width expands (0.5 ms), and the continuous phase flows around the droplets at a velocity higher than that of the droplet (1.0–1.5 ms). Consequently, a certain amount of the sample is transported (dispensed) to the continuous phase side, which enables the concentration control by the droplet injection frequency while enhancing the mixing due to the reduction of the diffusion length (2.0–3.5 ms).

**Effects of the diffusion coefficient and the confluence angle**

Effects of the diffusion coefficient and the confluence angle on the mixing characteristics are investigated with four different $\theta_m$ (30, 45, 60, and 90°) and two different solutions (dye and fluorescent particle solutions). Figure 3 shows the mixing near the confluence point with different $\theta_m$ while the width of the mixing channel $W_m$ = 200 μm and the droplet injection frequency $f$ = 15 Hz are kept constant. The images for the dye solutions are obtained under the bright-field condition, whereas those for the particle solutions are taken under the dark-field condition. Figures 3(a) and 3(b) reveal that the mixing is enhanced even with relatively low diffusion coefficient sample (typical diffusion coefficient for dye and 1-μm particles are $10^{-10}$ and $10^{-16}$ m$^2$ s$^{-1}$, respectively). However, the effects of the diffusion coefficient is not negligible when we evaluate the RMI. In general, mixing level of the dye solution is higher than that of the particle solutions because the diffusion drives the homogenization after the perturbation of the mixing front by the droplets. Moreover, Figs. 3(c) and 3(d) show that the decrease in the RMI at $x_m$ > 2.5 mm is only seen in the case of the dye solution. It suggests that the mixing enhancement by the droplet injection is caused only in a region near the confluence point and the diffusion is dominant in the downstream region. The dominance of the diffusion in the downstream region is also suggested by the fact that the inclination of the RMI slopes in $x_m > 2.5$ mm are similar to those for the "no droplet" cases. In addition, the RMI for the dye solution is lower (*i.e.*, higher mixing level) than that for the particle solution presumably because of the high mobility of the dye.

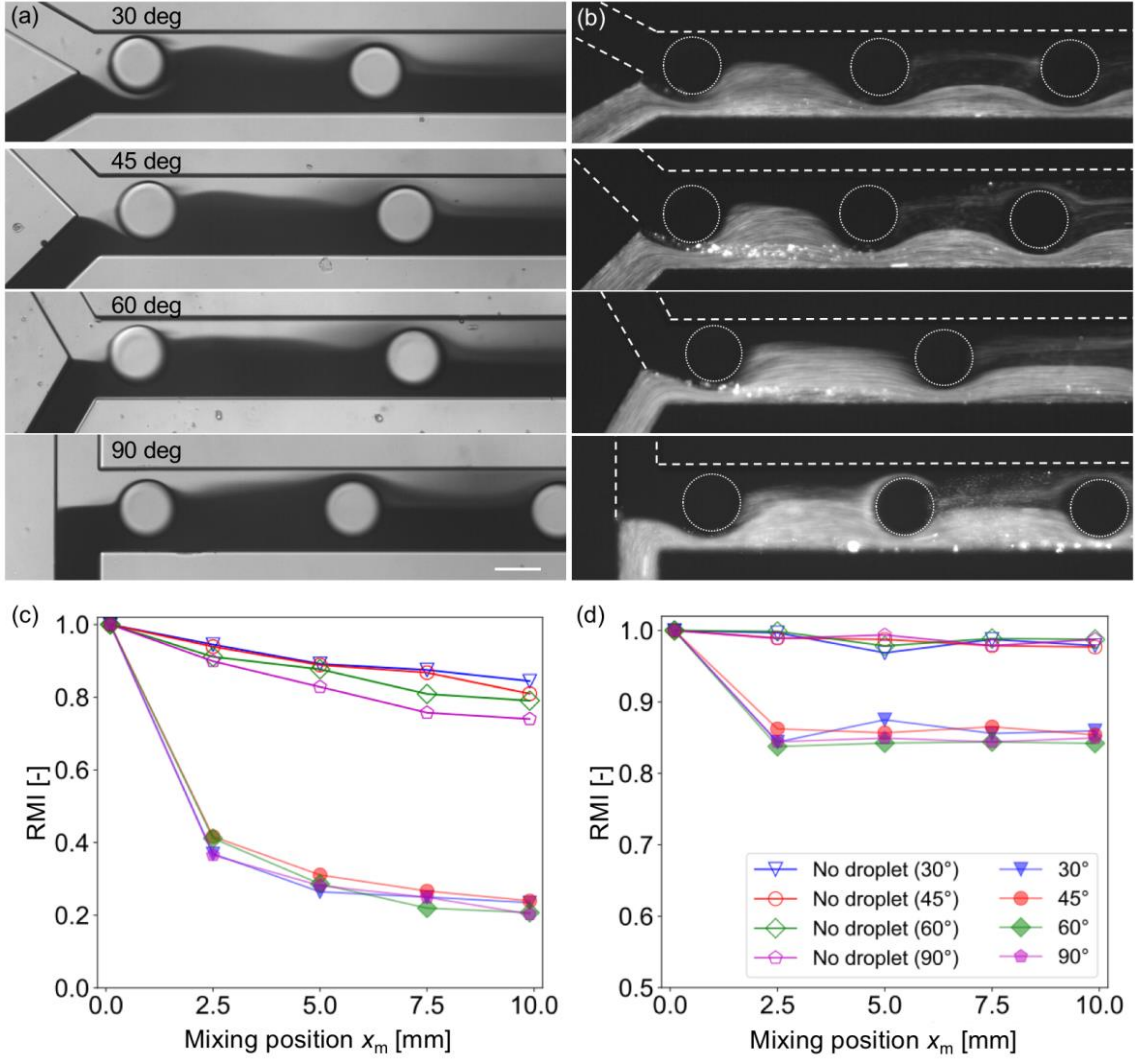

Fig. 3 Effects of the confluence angle and the diffusion coefficient on the mixing characteristics. Mixing of (a) the dye and (b) the fluorescent polystyrene particles (1 μm in diameter) in channels whose confluence angle $\theta_m$ are 30, 45, 60, and 90°. Scale bar indicates 100 μm. Relationships between the RMI and channel location for (c) the dye and (d) the particles indicate that the effect of the droplet injection is dominant over that of the confluence angle. Mixing enhancement by the droplet injetion for the particles (RMI = 0.85) is lower than that for the dye (RMI = 0.2) due to their large diffusion coefficient.

The effect of the confluence angle on the mixing enhancement is observed for the "no droplet" case as the difference in the RMI at $x_m = 10$ mm [Fig. 3(c)] in the dye solution. The diagram indicates that the mixing is enhanced as $\theta_m$ increases. This may be due to an increase in the influence of the change of the momentum on the flow field[39, 40]. Conversely, the angle effect is suppressed in the cases of the particle solution and of the enhanced mixing by the droplets. These results suggest that the effect of the droplet injection is significant and that of the confluence angle is negligible for low diffusion coefficient samples and for the droplet injection mixing.

**Effects of the droplet injection frequency**

Effects of the frequency of the droplets injected into the mixing channel is investigated. In this experiment, $W_m = 250$ μm and $Q_{sample} = (Q_{cont} + Q_{disp}) = 420$ μL h$^{-1}$ are treated as constants, while the injection frequency $f$ is changed by controlling the flow rate ratio $\varphi$. The frequency range is limited by the chip design and the working fluids to $f \leq 15$ Hz (see also: Fig. S2 in Supplementary

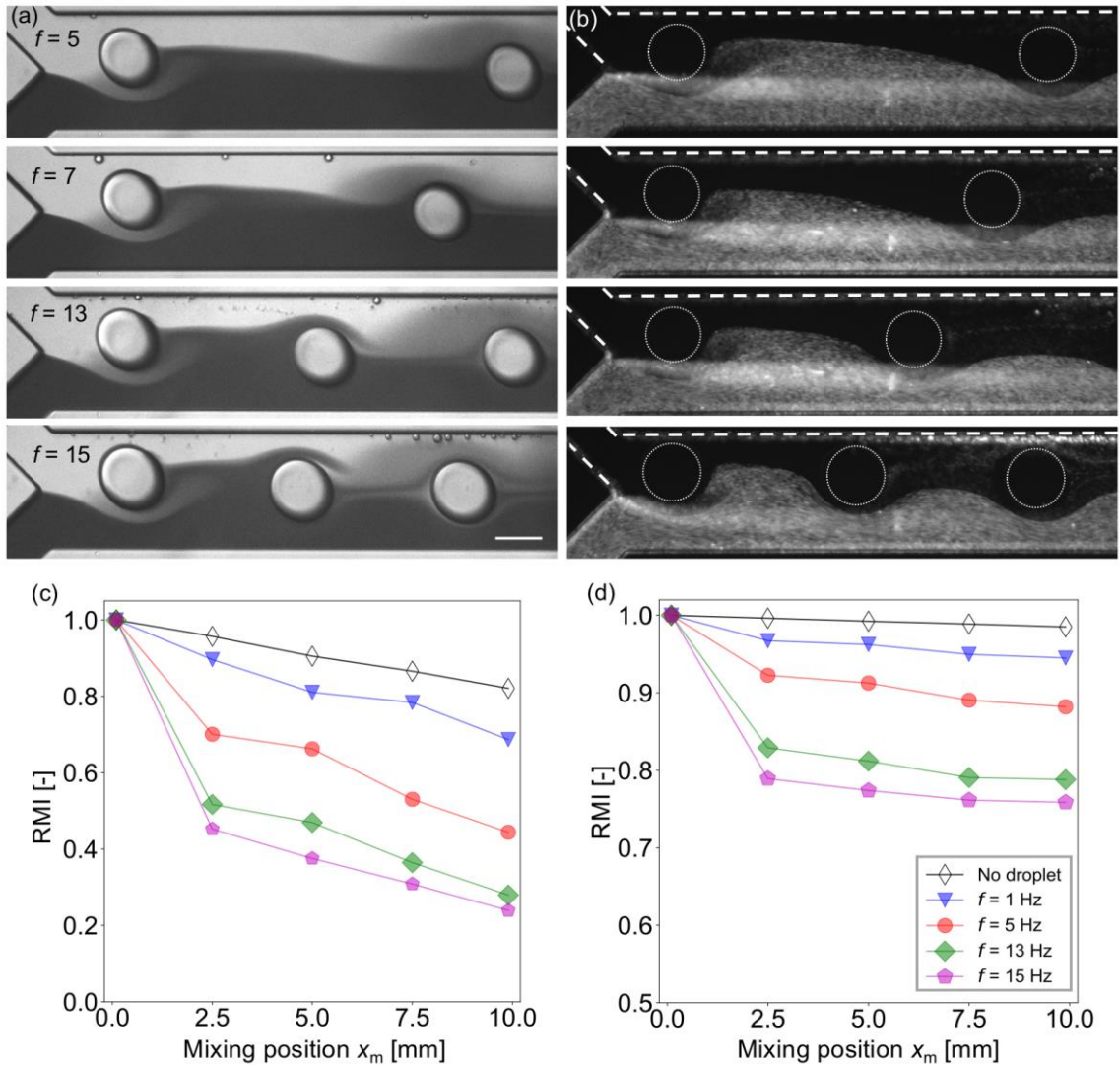

Fig. 4 Effects of the droplet injection frequency on the mixing characteristics. Mixing of (a) the dye and (b) the particles at $f$ = 5, 7, 13, and 15 Hz. Scale bar indicates 100 μm. (c) and (d) indicate the relationship between the RMI and the mixing position in the cases of the dye and the particles, respectively. The resutls indicate that the solution concentration is controllable by adjusting the injection frequency regardless of the solute diffusion coefficient. The edges of the droplets in (b) are enhanced by white circles only for the purpose of eye guide.

Information). For $f > 15$ Hz, the droplets are not generated and co-flows are formed[41]. Figures 4(a) and 4(b) show that the distance between the neighboring droplets decreases as the frequency increases. As it has been discussed in the previous section, the mixing enhancement by the droplet is the maximum near the confluence point. Figures 4(c) and 4(d) indicate that the mixing is more enhanced for higher frequency, whereas the inclination of the RMI for $x_m > 2.5$ mm is almost independent of $f$ for both the dye and the particle solutions.

**Effects of the droplet size**

Effects of the droplet occupancy in the width of the mixing channel is investigated under constant mean velocity and droplet diameter $D_d$ condition. Note that the difference in the velocity does not affect the mixing characteristics except for the diffusion effect in the downstream region (see Fig. S3 in Supplementary Information). Note also that the droplet occupancy, defined as $D_d / W_m$, over the unity ($D_d / W_m > 1$) indicates that the droplets are not spherical due to the confinement by the side walls. The experiment is carried out with five different $W_m$ ranging from 100 μm to 300 μm, in which the occupancy ranges from 0.3 to 1.2. Figures 5(a)–5(e) show the droplets injected into the mixing channel of different $W_m$. The images show that the perturbation occurred within the distance of

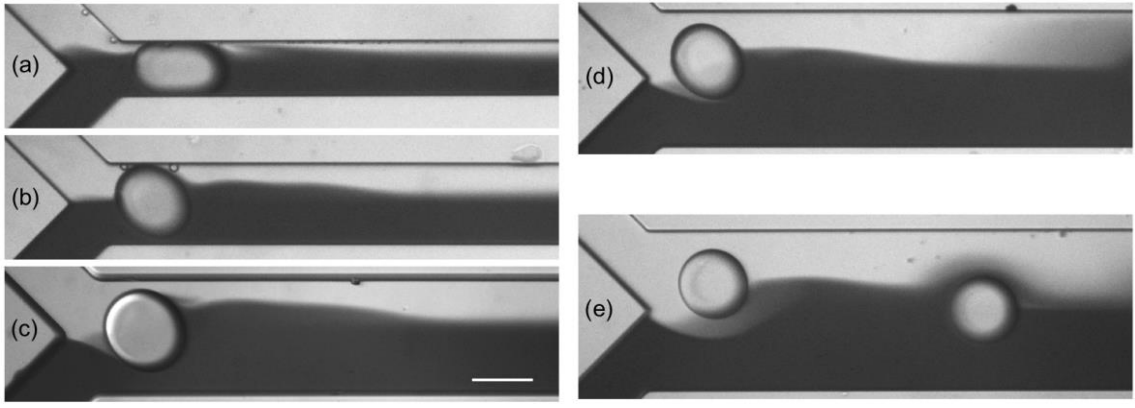

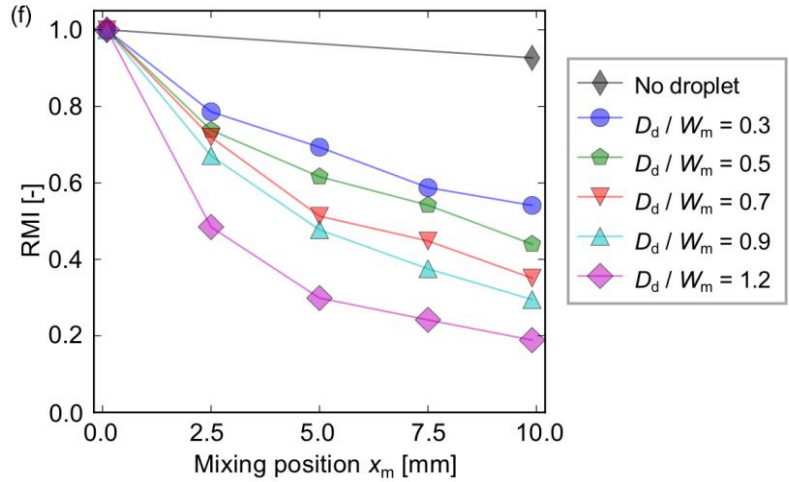

Fig. 5 Effects of the relative droplet size $D_d / W_m$ on the mixing characteristics. (a)–(e) Mixing for the cases of $D_d / W_m$ = 1.2, 0.9, 0.7, 0.5, and 0.3. (f) The relationship between the RMI and mixing position in a case of the dye. The results indicate that the resulting concentration depends on the relative droplet size. Scale bar indicates 100 μm.

~$D_d$ from the interface regardless of the channel geometry. The RMI curves [Fig. 5(f)] show that the mixing is highly enhanced in more confined case. It implies that high confinement is favorable for obtaining high mixing level.

**Discussion**

We developed a micromixing technique that can control the mixing level (*i.e.*, sample concentration) and demonstrated the mixing of sample (either dye or particle solutions) and buffer. A series of the experiment revealed that the mixing is enhanced by the droplet injection onto the sample–buffer interface whereby a certain amount of the sample is transferred to the buffer side and the area of the mixing front is increased. The mixing enhancement effect is the maximum at the inlet of the mixing channel, and the diffusion dominates the mixing at the downstream region. The concentration of the mixed solution is then spatially homogenized after being extracted in the outlet channel due to the circulation flow generated in the Taylor flow. These mixing characteristics are independent of the flow velocity in the mixing channel in our experiment range of 5–50 mm s$^{-1}$ (see Supplementary Information and Fig. S3). In the following part, we discuss the mass transport mechanism by the droplet injection in more detail to better understand the physics of the mixing and to gain accurate concentration controllability.

The successive images in Fig. 2(c) show that the buffer phase overlaps a droplet, which passed the confluence point, through gaps between the droplet and the channel walls. As a result, the flow rate at both sides are ~$Q/2$ and ~$3Q/2$, respectively, where $Q$ denotes the total flow rate of the three phases. The separation generates a fluctuation of the mixing front because this separation of the flow happens periodically. Furthermore, the droplet motion in the cross-sectional direction significantly enhances the mixing. In our system, the droplet Reynolds number $Re_d$ (= $\rho u D_d \mu^{-1}$, where $\rho$, $u$, and $\mu$ denote liquid density, mean velocity in the mixing channel, and liquid viscosity, respectively) ranges from 0.5 to 5: in this range, the inertial lift force exerts on the droplets and they migrate in the cross-sectional direction towards the channel center[42, 43]. This migration generates a substantial spanwise mass transfer near the confluence point as it can be seen at 1.5–3.5 ms in Fig. 2(c). Consequently, a certain amount (roughly estimated as a half volume of the droplet) of the sample is transferred toward the buffer phase and it flows in the upper side region as the droplets settle down at the center of the channel. It implies that the mixing level can be directly controlled by changing the volume of the sum of the droplets in unit volume (a segmented volume which contains one droplet). This trend is indeed observed in Figs. 4(c) and 4(d) where the mixing level increases with the droplet frequency under the constant flow rate condition.

Unlike other droplet/bubble-based mixing technique, the developed technique exploits the flow around the droplets to control the mixing. Postek *et al.*[34] developed a droplet-based dilutor in which the dilution ratio is controlled by a number of repetitions of the sample–buffer droplet coalescence. This technique is suitable for producing a small amount of sample because the device can control the droplet concentration precisely as well as the droplet volume. However, it is not suitable for mass production because of its time-consuming nature for requiring the repetitive operation to achieve a particular concentration. Mao *et al.*[25] generated chaotic flows by introducing bubbles in mixing chambers. The chaotic flows are the result of the random bubble motions inside the chambers so that strong convective flows between adjacent bubbles are induced. The technique is highly effective for the homogenization, but it is not designed to control the mixing level. In contrast to these techniques, the droplets have two different roles in the developed technique: at the confluence point, they generate the convective flow to extract the sample, whereas they enhance the mixing in the outlet channel by shortening the diffusion length[24]. These two roles enable the custom-made mixing by "picking-up" and "stirring" the solution in one device and make the concentration-adjustment operation straightforward. Figure 6 shows evolutions of the sample concentration (normalized by the highest concentration of each case) as a function of the droplet volume fraction. It clearly indicates that the concentration is a linear function of the volume fraction for both the dye and the particle solutions. These results suggest that the convection due to the droplet impingement is the dominant mechanism of the mixing and that our technique can produce a large amount of arbitrary-concentration-controlled sample solution with a simple operation.

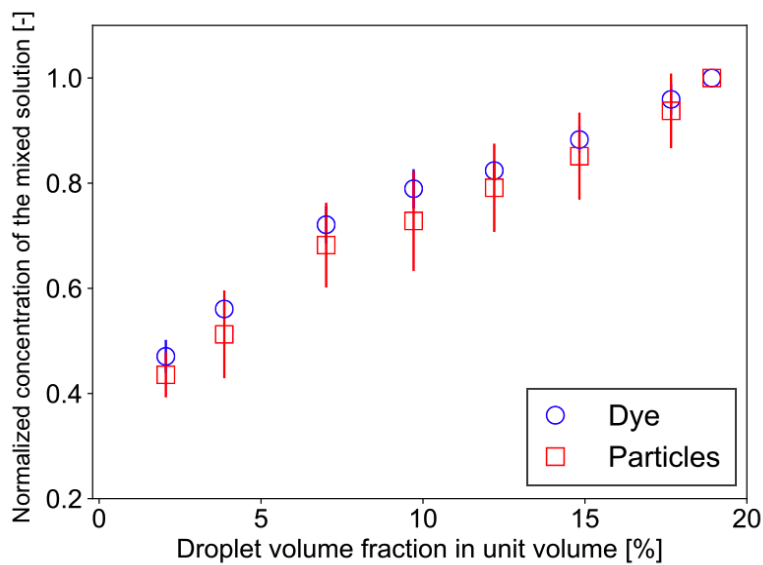

Fig. 6 Evolution of the sample concentration as a function of the droplet volume fraction. Horizonal and vertical axes indicate the droplet volume fraction in a unit volume and the concentration of the mixed sample (normalized by the highest concentration in the series), respectively.

## Methods

### Device fabrication

The PDMS (polydimethylsiloxane) micromixer is fabricated by the standard soft lithography technique[44]. First, the PDMS is mixed with a curing agent at a mixing rate of 10:1 and the mixed solution is degassed. The degassed solution is then poured into a mold on a silicon wafer made of SU-8 and cured at 80°C for 60 minutes. The cured PDMS is removed from the mold, holes are punched for the tubing, and Teflon tubes are connected to the holes. Finally, the PDMS and a glass substrate are bonded after an oxygen plasma treatment for improving their hydrophilicity and bonding characteristics.

### Experimental setup

The flow inside the device is recorded with 2000 fps from downward direction using a high-speed camera attached on an optical microscope (see also: Fig. S1 in Supplementary Information). We use two different lighting: a bright-field observation with white light emitted by an LED light is carried out for the measurement of the dye solution, whereas a dark-field observation with 505 nm incident light is carried out for the measurement of the fluorescent particle solution. DI water and oleic acid are chosen as the continuous phase and the dispersed phase, respectively. We used dye solution (mixture of the brilliant blue FCF and DI water) and particle solution [mixture of fluorescent polystyrene particle (1 μm in diameter) and DI water] as the sample solution. The working fluids are infused by syringe pumps at constant flow rate and the mixture of them are extracted from two outlets connected to an outlet reservoir.

### Control of the droplet injection frequency

The droplet injection frequency (*i.e.*, droplet generation frequency at the T-junction) is stably varied by controlling the flow rate ratio of the continuous and the dispersed phases. In our system, the flow rate ratio $\varphi$ can be varied from 0.01 to 0.20 (see Fig. S2 in Supplementary Information). In the case of $\varphi > 0.20$, co-flows are formed. This is mainly because we did not add any surfactant.

### Quantification methods for the mixing enhancement

The mixing level is quantitatively evaluated by calculating the RMI (relative mixing index)[38] by substituting a standard deviation of a light intensity distribution in a specified region inside a channel into equation (1). $\sigma$, $I_i$, $I_{0i}$, $I_m$, and $N$ in euqation (1) denote the standard deviation of the intensity, the intensity of each image pixel, the intensity of an image pixel before the mixing, and the average intensity of the analysis field, respectively. The RMI is exploited as an indicator of the mixing level, where RMI = 1 and RMI = 0 indicate complete separation and complete mixing, respectively. In this study, the RMI in the mixing channels at five different point (0.0, 2.5, 5.0, 7.5, and 10.0 mm from the confluence point) are measured.

$$RMI = \frac{\sigma}{\sigma_0} = \frac{\sqrt{\frac{1}{N}\sum_{i=1}^{N}(I_i - I_m)^2}}{\sqrt{\frac{1}{N}\sum_{i=1}^{N}(I_{0i} - I_m)^2}} \qquad (1)$$

## Acknowledgements

This work was financially supported by JSPS KAKENHI Grant Number 16K18033.

## Author contributions statement

K.Y., R.S., and M.M. conceived the experiments, R.S. conducted the experiments, K.Y., R.S., and M.M. analyzed the results, K.Y. established the physical model, R.S. and K.Y. prepared Figures and the graphs, R.S. and K.Y. wrote the manuscript. All authors reviewed the manuscript.

## Additional information

**Supplementary Information** accompanies this paper

**Competing financial interests:** The authors declare no competing financial interests.

# Concentration-adjustable micromixer using droplet injection into a microchannel

Ryosuke Sakurai[1,+], Ken Yamamoto[1,2,*,+], Masahiro Motosuke[1,2]

[1]Department of Mechanical Engineering, Tokyo University of Science, 6-3-1 Niijuku, Katsushika-ku, Tokyo 125-8585, Japan

[2]Research Institute for Science and Technology, Tokyo University of Science, 6-3-1 Niijuku, Katsushika-ku, Tokyo 125-8585, Japan

*Corresponding author, yam@rs.tus.ac.jp

+these authors contributed equally to this work

## Supplementary Information

**Experimental setup**

The flow inside the device is recorded with 2000 fps from downward direction using a high-speed camera attached on an optical microscope (Fig. S1). We use two different lighting: a bright-field observation with white light emitted by an LED light is carried out for the measurement of the dye solution, whereas a dark-field observation with 505 nm incident light is carried out for the measurement of the fluorescent particle solution. DI water and oleic acid are chosen as the continuous phase and dispersed phase, respectively. We used dye solution (mixture of the brilliant blue FCF and DI water) and particle solution [mixture of fluorescent polystyrene particle (1 μm in diameter) and DI water] as the sample solution. The diffusion coefficient of the dye and the particles in water are estimated as $10^{-10}$ m$^2$ s$^{-1}$ and $10^{-16}$ m$^2$ s$^{-1}$, respectively. The fluorescent particles are excited by the 505 nm light and emit 515 nm (peak wavelength) light. We did not add any surfactant in this experiment. The emitted light is extracted by the camera through a dichroic mirror and a bandpass filter. The working fluids are infused by syringe pumps at constant flow rate and the mixture of them are extracted from two outlets connected to an outlet reservoir.

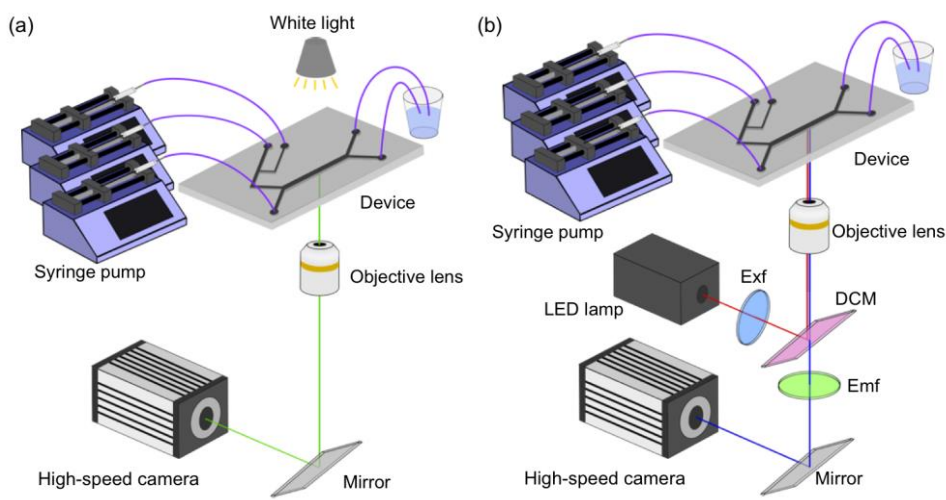

Fig. S1 Experimental setup for mixing of (a) the dye and (b) the particle solutions. (a) An LED white light is used for the bright-field imaging in the case of the dye mixing, whereas (b) the dark-field imagin with the incident light of the wavelength of 505 nm is irradiated in the case of the particle mixing. The particles emit 515 nm light.

**Control of the droplet injection frequency**

The droplet injection frequency (*i.e.*, droplet generation frequency at the T-junction) is stably varied by controlling the flow rate ratio of the continuous and the dispersed phases. In our system, the flow rate ratio $\varphi$ can be varied from 0.01 to 0.20 (Fig. S2). In the case of $\varphi > 0.20$, co-flows are formed. This is mainly because we did not add any surfactant.

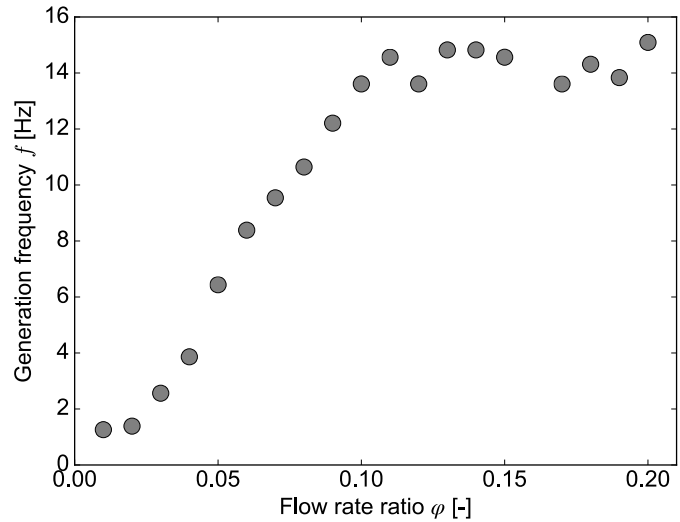

Fig. S2 The droplet generation frequency as a function of the flow rate ratio $\varphi$.

**Effects of the velocity**

The effect of the velocity is investigated in a mixing channel of $W_m = 200$ µm. Figure S3 shows the mixing with three different mean velocity in the mixing channel ($u = 5.0, 20.0,$ and $50.0$ mm s$^{-1}$). The droplet Reynolds number and the Peclet number (Pe = $uL / D$, where $u$, $L$, and $D$ are the characteristic velocity, characteristic length, and the diffusion coefficient of the sample, respectively) for the three cases are Re$_d$ = 0.5, 2, and 5, and Pe = $1 \times 10^4$, $4 \times 10^4$, and $1 \times 10^5$, respectively.

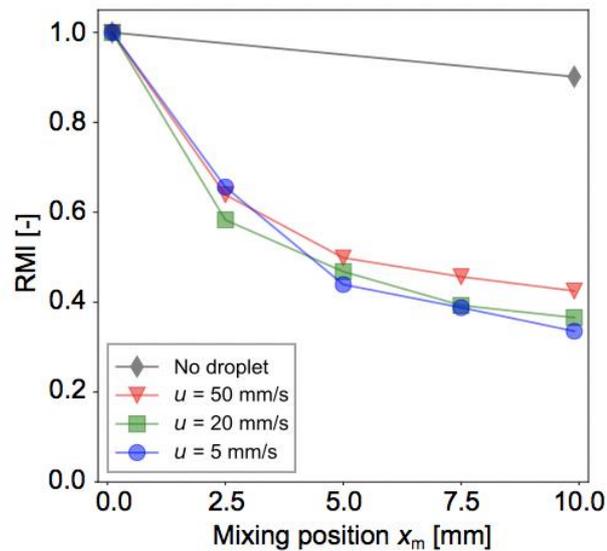

Fig. S3 Measured RMI for three different mean velocities ($u = 5, 20,$ and $50$ mm s$^{-1}$). The result indicates that the resulting concentration does not depend on the velocity in the mixing channel.

**Mixing of particles in an expanding channel**

A mixing experiment in an expanding channel (Fig. S4) is conducted. The channel width expands from 200 μm (at the confluence point) to 600 μm (at the end of the mixing channel, 10 mm downstream of the confluence point) and the sample and the buffer are introduced in the channel with $Q_{sample} = (Q_{cont} + Q_{disp}) = 420$ μL h$^{-1}$. In contrast to Fig. S4(a), the mixing of the particle solution is significantly enhanced by introducing the droplets [Fig. S4(b)]. In this channel, the mixing is more enhanced than the case in the constant-width channel, because the migration of the droplets in the cross-sectional direction increases as the channel expands whereby the droplets touch each other and generate a chaotic advection. Although some droplet coalescence are observed because of the absence of surfactants, it is observed that the mixing enhancement effect is significant regardless of the coalescence. The channel is considered to be suitable to achieve homogeneity rather than the concentration control.

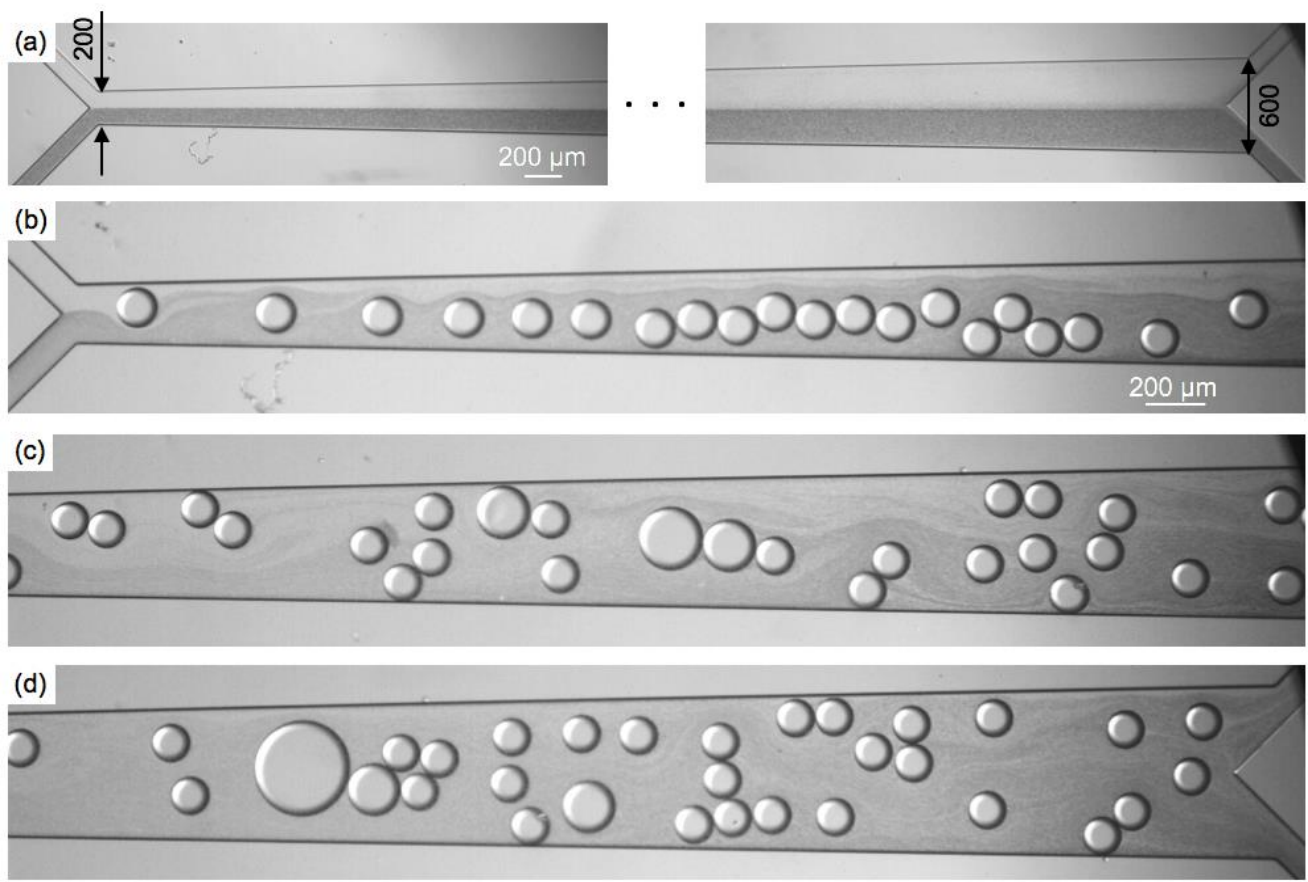

Fig. S4 Mixing of the particles (1 μm in diameter) in an expanding channel in which $W_m$ gradually changes from 200 to 600 μm. (a) Diffusive mixing is dominant on the interface of the continuous phase and the sample in the case of no droplet injection. (b)–(d) A series of the injected droplets collapse each other due to the channel-width expansion and the mixing efficiency is enhanced by the chaotic advection induced by the droplets.